# Coarse-grained model for spring friction study of micron-scale iron by smoothed particle hydrodynamics

LE VAN SANG[1(b)], AKIHIKO YANO[2], SHUJI FUJII[2], NATSUKO SUGIMURA[1,3],

HITOSHI WASHIZU[1(a)]

[1]Graduate School of Simulation Studies-University of Hyogo

[2]Mitsubishi Heavy Industries, Ltd.

[3]Faculty of Engineering-Tokyo City University

[(a)]Email: h@washizu.org

[(b)]Email: levansang82@gmail.com

**Abstract** – The paper constructs a coarse-grained model to investigate dry sliding friction of the body-centered-cubic Fe micron-scale system by smoothed particle hydrodynamics simulations and examines influences of the spring force on the characters of friction. The $N_{atom} = 864 \times 10^{12}$ atoms Fe system is coarse-grained into the two different simple-cubic particle systems, one of 432000 and the other of 16000 particles. From the detection of stick-slip motion, friction coefficient, dependence of friction coefficient on isotropy or anisotropy of the spring force and externally applied normal load, we find that the coarse-grained model is a reasonable modeling process for study of friction of the Fe system and the anisotropic behavior presents better friction of the system than the isotropic one.

**Introduction.** – Coarse-graining (CG) has become one of the most noted choices in simulations of micron-scale systems. The number of degree of freedom decreases many times in a CG system compared to that of a corresponding atomistic system; therefore, one has to solve only a few equations of motion for a CG system. Furthermore, since mass of a CG particle is much larger than that of an atom, a simulation time step can be chosen a longer one leading to that a CG system can be simulated in a long time. In theoretical approach, Kinjo et al. derived the equation of motion of the coarse-grained particles by the projection operator method [1]. There were a few simulation studies of mechanical properties of CG solid materials. Simulations of Rudd et al. indicated that CG molecular dynamics provided a better description of the elastic waves than that provided by finite element modeling and the elastic wave scattering was more benign in CG molecular dynamics than finite element modeling for the small periodic CG solid systems [2,3]. Kobayashi et al. supposed a CG model named the recursive CG particle method to investigate large realistic systems [4]. Although Kobayashi et al. well applied their model to investigate deformation of the Ar and Al nanorods, they simulated the two or three dimensional systems of sub-micron size. One needs to investigate a micron-scale system to be able to observe mechanical properties of devices. Nearby, Yang et al. [5] coarse-grained a Cu substrate by estimation of its new face-centered-cubic (fcc) lattice constant based on the formula of the nearest neighbor distance of the fcc Lennard-Jones substrate established by Stevens [6]. Limit of this coarse-graining is only one permitted CG time leading to that the particle number of a system does not decrease much. This is not reasonable for considering large systems. Simulation results of the CG Cu systems exhibited a good agreement of change of stress between a CG system and an atomistic one [7]. However, due to choosing surfaces of particle oriented along {111} close-packed planes,

Xiong et al. calculated mass of each particle in approximation and designed a complex CG method. A CG model should also fully meet simplicity of modeling process.

At atomistic level, we have been interested in the origin of kinetic friction by calculating phonon dissipation of bulk atoms [8-10]. Our recent work carried out sliding friction of the GC graphene sheets [11]. Thermal escape motion resulting in low friction mechanism in the particle model [11] is similar to that confirmed in atomistic systems. Our present work presents a simple CG method of three-dimensional crystal system where mass of particle is taken exactly. In this method, a body-centered-cubic (bcc) crystal system is first coarse-grained into a simple-cubic (sc) crystal system. The obtained sc system is then coarse-grained into a larger sc one. This CG method is completely based on symmetry of lattice structures and can easy be modeled for a large system. The GC model is utilized to investigate sliding friction of a Fe micron-scale system by smoothed particle hydrodynamics (SPH) simulations. Up to now, we have only found a study of friction of a Fe nano-scale system at sliding velocities of 100 – 350 m/s [12] while most studies have concerned about that of Fe alloys [13,14]. In recent years, research groups have reported friction or tribology of various materials at high sliding velocities, 100 m/s [15], 400 m/s [16] and 120 m/s [17]. Along with these studies, the present work investigates friction of the Fe system at a sliding velocity of 100 m/s. Moreover, effects of spring force on friction are also carefully discussed via spring constant and isotropy or anisotropy of spring force.

**The CG model and SPH simulation.** – Coarse-graining is proceeded in two steps: firstly, each unit cell of the initially bcc atomistic Fe system is coarse-grained into one particle that has mass of $2m$ ($m = 55.845$ g/mol) and is positioned at the center of the unit cell. This coarse-graining exchanges the atomistic system (fig. 1(a)) to the sc particle one (fig. 1(b)) having a lattice constant of $a_{BCC} = 2.85$ Å; finally, the obtained sc particle system is continuously coarse-grained into a larger sc particle one by grouping a cubic region of $N_l^3$ unit cells ($N_l$ in each direction x, y or z) into one particle. The final sc particle system (fig. 1(c)) has the following characters: mass of each particle $M_{CG} = 2N_l^3 m$, a lattice constant $a_{CG} = N_l a_{BCC}$ and the number of particles $N_{CG} = \dfrac{N_{atom}}{2N_l^3}$.

Due to its well-known calculations, the SPH method is briefly presented by the governing equations in the present work. The time-evolution of the density $\rho_i$, velocity $v_i^\alpha$ and internal energy $u_i$ of the $i$th particle is described by the following equations

$$\dfrac{d\rho_i}{dt} = \sum_{j=1}^{N} m_j (\vec{v}_j - \vec{v}_i) \vec{\nabla}_i W(\vec{r}_i - \vec{r}_j, h), \tag{1}$$

$$\dfrac{dv_i^\alpha}{dt} = \sum_{j=1}^{N} m_j \left( \dfrac{\sigma_i^{\alpha\beta}}{\rho_i^2} + \dfrac{\sigma_j^{\alpha\beta}}{\rho_j^2} + \Pi_{ij} \right) \nabla_i^\beta W(\vec{r}_i - \vec{r}_j, h), \tag{2}$$

$$\dfrac{du_i}{dt} = \dfrac{1}{2} \sum_{j=1}^{N} m_j \left( \dfrac{\sigma_i^{\alpha\beta}}{\rho_i^2} + \dfrac{\sigma_j^{\alpha\beta}}{\rho_j^2} + \Pi_{ij} \right) (v_j^\alpha - v_i^\alpha) \nabla_i^\beta W(\vec{r}_i - \vec{r}_j, h), \tag{3}$$

where $\alpha$ or $\beta \equiv x, y, z$; $m$ and $\vec{r}$ are mass and position vector, respectively; the super Gauss kernel function

$$W(\vec{r}_i - \vec{r}_j, h) = \dfrac{1}{\pi^{3/2} h^3} \left( \dfrac{5}{2} - r^2 \right) e^{-\dfrac{r^2}{h^2}}, \tag{4}$$

here $r = |\vec{r}_i - \vec{r}_j|$, $h = h_i = \eta \left( \dfrac{m_i}{\rho_i} \right)^{1/3}$ is smoothed length of the $i$th particle and $\eta$ is a parameter; the artificial viscosity function

$$\Pi_{ij} = \begin{cases} \dfrac{-A\bar{c}_{ij}\mu_{ij} + B\mu_{ij}^2}{\bar{\rho}_{ij}} & \vec{v}_{ij}.\vec{r}_{ij} \leq 0 \\ 0 & \vec{v}_{ij}.\vec{r}_{ij} > 0, \end{cases} \tag{5}$$

here $\vec{v}_{ij} = \vec{v}_i - \vec{v}_j$, $\vec{r}_{ij} = \vec{r}_i - \vec{r}_j$, $A$ and $B$ are parameters, $\bar{c}_{ij} = \frac{1}{2}(c_i + c_j)$, $c = \sqrt{\frac{h_{cr} p}{\rho}}$ is sound speed of the particle, $\bar{\rho}_{ij} = \frac{1}{2}(\rho_i + \rho_j)$, $\mu_{ij} = \frac{h \vec{v}_{ij} \cdot \vec{r}_{ij}}{r^2 + \cdot h^2}$, $p = (h_{cr} - 1)\rho u$ is pressure of the particle and $h_{cr}$ is a parameter; and the stress tensor

$$\sigma_i^{\alpha\beta} = -p_i \delta^{\alpha\beta} + S_i^{\alpha\beta}, \tag{6}$$

here $\delta^{\alpha\beta}$ is the Kronecker symbol, $S_i^{\alpha\beta}$ is the deviatoric stress calculated from the equation

$$\frac{dS_i^{\alpha\beta}}{dt} = 2\mu \left( \dot{\varepsilon}_i^{\alpha\beta} - \frac{1}{3}\delta^{\alpha\beta} \dot{\varepsilon}_i^{\gamma\gamma} \right) + S_i^{\alpha\gamma} R_i^{\beta\gamma} + S_i^{\gamma\beta} R_i^{\alpha\gamma}, \tag{7}$$

in which $\gamma \equiv x, y, z$ and $\mu$ is the shear modulus of materials, $\dot{\varepsilon}_i^{\alpha\beta}$ is the tensor of the rate of deformations defined by

$$\dot{\varepsilon}_i^{\alpha\beta} = \frac{1}{2} \sum_{j=1}^{N} \frac{m_j}{\rho_j} \left[ \left(v_j^\alpha - v_i^\alpha\right) \nabla_i^\beta W(\vec{r}_{ij}, h) + \left(v_j^\beta - v_i^\beta\right) \nabla_i^\alpha W(\vec{r}_{ij}, h) \right], \tag{8}$$

and $R_i^{\alpha\beta}$ is the tensor of stress rotation defined by

$$R_i^{\alpha\beta} = \frac{1}{2} \sum_{j=1}^{N} \frac{m_j}{\rho_j} \left[ \left(v_j^\alpha - v_i^\alpha\right) \nabla_i^\beta W(\vec{r}_{ij}, h) - \left(v_j^\beta - v_i^\beta\right) \nabla_i^\alpha W(\vec{r}_{ij}, h) \right]. \tag{9}$$

We consider adding the dissipation force on each particle to compensate energy dissipation caused by friction during the sliding

$$F_{dis, i} = \begin{cases} -m_i \gamma_{dis} \left(v_i^x - V_{dis}\right) & \text{the x-direction} \\ -m_i \gamma_{dis} v_i^y & \text{the y-direction} \\ -m_i \gamma_{dis} v_i^z & \text{the z-direction,} \end{cases} \tag{10}$$

where $\gamma_{dis}$ is a parameter of the model and $V_{dis} = 0$ for particles of the substrate and $V_{dis} = V$, which is a constant sliding velocity of the slider in x-direction, for particles of the slider. By using the Prandtl-Tomlinson model, we also add a spring force on each particle of the slider

$$F_{spr,i} = \begin{cases} K(x_{0,i} + Vt - x_i) & \text{the x-direction} \\ K(y_{0,i} - y_i) & \text{the y-direction} \\ K(z_{0,i} - z_i) & \text{the z-direction,} \end{cases} \qquad (11)$$

where $K$ is a spring constant, $t$ is a sliding time, $x_{0,i}$, $y_{0,i}$ and $z_{0,i}$ are the equilibrium/initial coordinates of the $i$th particle in the x, y and z-directions, respectively. Interaction between the slider and the substrate is presented by interaction between particles of the lowest layer of the slider and particles of the highest layer of the substrate. Two particles, one of each layer, interact with each other via the following spring force

$$\vec{F}_{int,ij} = \begin{cases} -k_\alpha (r-h)\dfrac{\vec{r}_{ij}}{r} & 0 < r < h \\ 0 & r > h, \end{cases} \qquad (12)$$

where $k_\alpha$ is a spring constant that is considered isotropy or anisotropy in the present study. The friction force $F_{fri}$, the normal force $F_{nor}$ and the friction coefficient $\mu_{cof}$ are defined as follows

$$F_{fri} = \sum_{i=1}^{N_f} \left( F_{spr,i}^x + F_{int,ij}^x \right), \qquad (13)$$

$$F_{nor} = \sum_{i=1}^{N_f} \left( F_{spr,i}^z + F_{int,ij}^z \right), \qquad (14)$$

$$\mu_{cof} = \frac{F_{fri}}{F_{nor}}, \qquad (15)$$

where $N_f$ is the number of the particles of the lowest layer of the slider, $F^x$ and $F^z$ are the force components in the x and z-directions, respectively.

The $N_{atom} = 864 \times 10^{12}$ atoms bcc Fe system including the slider of $17.10 \times 17.10 \times 8.55$ (μm³) and the substrate of $51.30 \times 17.10 \times 8.55$ (μm³) is modeled into the two different CG systems, one of $N_l = 1000$ with $N_{CG} = 432000$ (fig. 1(d)) and the other of $N_l = 3000$ with $N_{CG} = 16000$ (not shown). The initial distance between the two objects is equal to $a_{CG}$ of each CG system. The lowest two layers of the substrate are fixed during the simulations. The parameters are used in the simulations: $\mu = 52.5$ GPa [18], $\eta = 1.2$, $A = B = 0.1$, $h_{cr} = 1.4$, $\cdot = 0.01$, $\gamma_{dis} = 10^3$ 1/s, $\rho = 7.86$ g/cm³, $V = 100$ m/s, a time step $dt = 285$ ps and $K = 0.051$ nN/nm that was used as a spring constant of the conventional cantilevers in study of fresh iron particles [19]. We modify the FDPS open source developed by Iwasawa et. al. [20] to create our simulation program.

**Results and discussions.** – Figure 2 shows the sliding time-dependence of the friction force, normal force and friction coefficient (CoF) of the systems in the behavior of load of 0.1 μN and $k_z = k_y = 0.1 k_x$ with $k_x = 0.2K$. A regular occurrence of the stick-slip motion is clearly observed from the curves. In spite of the sliding at the high velocity, the stick-slip motion still appears because the chosen values of the spring constants always guarantee a steady state of the particle layers at the interface and the objects are stably maintained by the damping parameter $\gamma_{dis}$. Sang et al. found a stick-slip type of instantaneous friction force in sliding of a support on a hard substrate [21]. The hard model in our considerations can be seen from a well-repeated periodicity of the saw-tooth shape. Note that the stick-slip friction abruptly disappeared above some critical driving velocity at which the melted or disordered film was no longer able to reorder during the motion [22]. We find that distance between the nearest peaks of each curve is exactly equal to the lattice constant of each system, $a_{CG} = 0.855$ μm for the 16000 particles system and $a_{CG} = 0.285$ μm for the 432000 particles system. This result is consistent with observations of lattice constant of various materials in the following experimental studies of sliding friction. Morita et. al. obtained a lattice constant of 3.16 Å for $MoS_2$ from measurements of friction force parallel to a $MoS_2$ surface acting on a $Si_3N_4$ tip [23]. Lattice constant of mica was detected by lateral force image of its (0001) surface [24]. Stick-slip motion with the periodicity of the KF unit cell was seen from lateral force image of KF (001) cleaved and imaged in UHV with a silicon nitride tip [25]. This scenario was also found for NaCl from measurements of the lateral force of a Si AFM tip sliding forward and backward in (100) direction over the NaCl (001) surface [26]. The two systems exhibit close to each other in oscillation amplitude and mean value of friction coefficient. This indicates that the CG model can be a reasonable process. In addition, a (mean) friction coefficient of ~ 0.3 monitored from the two curves (fig. 2) is in accordance with that of Fe as a dry sliding friction coefficient of about 0.15 – 0.40 under pressure of 9 – 45 kg/cm$^2$ [27] or 0.18 – 0.65 in sliding distance of 0 – 30 m [28]. These obtained results indicate that this CG model can be used to investigate sliding friction of Fe micron-scale system with the above simulation condition.

Figure 3 displays the friction coefficient dependent on value of $k_\alpha$ in the behaviors of load of 0.1 μN, $k_z = k_y = 0.1 k_x$ with $k_x = K$ (fig. 3(a)) and $k_x = 0.1 K$ (fig. 3(b)). The friction coefficient increases with decreasing the spring constant $k_\alpha$, around 0.058, 0.3 (the above discussion) and 0.565 corresponding to $k_x = K$, $0.2 K$ and $0.1 K$. Friction force is strong dependent on $k_x$ as seen in Eq. (13) while normal force is significant dependent on both $k_z$ and applied load; therefore, change of $k_x$ will strongly result in friction coefficient of the system. The value of 0.058 is much smaller than that reported in the previous studies [27,28], whereas the value of 0.565 is larger than a static friction coefficient (0.51) of Fe sliding on Fe in dry or unlubricated condition [29]. Limits of these considered cases can also be seen from strong oscillation (fig. 3(a)) and instability (fig. 3(b)) of the friction coefficient. However, regularity of the stick-slip motion is observed in all the simulations, seeming to be not dependent on value of the spring constant $k_\alpha$. Figure 4 depicts effects of isotropy or anisotropy of the spring interaction force $\vec{F}_{int,ij}$ on the friction coefficient in the behavior of load of 0.1 µN, $k_z = k_y$ and $k_x = 0.2K$. The friction coefficient grows as the anisotropy increases or the ratio $k_z/k_x$ decreases. This result is because that by fixing $k_x$ and increasing $k_z$ the friction force does not almost change while the normal force increases. Lemul et. al. reported that considering isotropy of materials in numerical models would lead minor differences between experimental and numerical results for friction coefficient [30]. Our result shows that the friction coefficient illustrates a small difference between the isotropic and anisotropic behaviors when the anisotropy is low, $k_z/k_x \geq 0.6$; however, it has an abrupt growth as the anisotropy is high, for example a $k_z/k_x$ drop from 0.2 to 0.1. The friction coefficient reaches to the experimental value [27,28] in the most anisotropic behavior $k_z/k_x = 0.1$. This result indicates that the isotropic or anisotropic consideration of the spring interfacial force strongly influences on friction coefficient of a system and supposes that the anisotropy should be carefully detected during collection of friction coefficient. In the past, most studies utilized spring force to monitor stick-slip configuration of friction force, whereas there were few studies of friction coefficient by using this force. Maveyraud et al. exhibited friction coefficient of CG solid rocks by using spring friction force; however, they did not mention effects of isotropy or anisotropy of this force on friction coefficient [31]. It is also important to note that there is a very good agreement of change of the friction coefficient between the two systems.

Figure 5 displays a decrease of the friction coefficient with an increase of the externally applied normal load in both the isotropic and anisotropic behaviors of the spring force with the simulation condition $k_x = 0.2K$ and $k_z = k_y$. Experimental studies have reported this state [32-37]. Cause of the decrease was explained by growing roughening of interface and forming a large quantity of wear debris [32]. The friction coefficient of the a-C:H coating and highly oriented pyrolytic graphite sample corresponding to the steady state reduces with increasing normal load because of elastoplastic deformation, wear, materials transfer and an increased real area of contact [33]. The slip of local precursors prior to the onset of bulk sliding led to the decrease of the macroscopic static friction coefficient [36]. Increased roughening and small deformations of the contacting surfaces play a vital role in our observations due to the surveyed steady interface. The curve approaching a linear one can be seen in the isotropic behavior $k_z/k_x = 1.0$ or low anisotropy $k_z/k_x = 0.5$. However, a nonlinear dependence of the friction coefficient via the load becomes more explicit in the high anisotropy $k_z/k_x = 0.1$. We find that the change of the curve, increment of the friction coefficient goes down with increasing the load, is similar to that found in experimental studies of dry or lubricated sliding friction of various materials [33,34,36,37]. The friction coefficient is in consistent with that of the experimental reports [27,28] in the range of the surveyed load in the case $k_z/k_x = 0.1$. The results also show that the two system are in accordance with each other in the applied load dependence of friction coefficient.

**Conclusions.** – This study uses smoothed particle hydrodynamics simulations to investigate sliding friction of the Fe micron-scale system by the CG model established by us and the spring friction force. A good agreement of the obtained results of the two particle systems demonstrates that the CG model is a reasonable modeling process. The spring constant and the isotropy or anisotroy of the spring friction force cause strong influences on the monitored friction quantities. In the anisotropic case with $k_z/k_x = 0.1$ and $k_x = 0.2K$, the CG model well presents the sliding friction characters of the Fe micron-scale system including the collection of the friction coefficient and the changes of the friction force and coefficient via the sliding time or the externally applied normal load.

**REFERENCES**

[1] KINJO T. and HYODO S., *Phys. Rev. E*, **75** (2007) 051109.

[2] RUDD R. E. and BROUGHTON J. Q., *Phys. Rev. B*, **58** (1998) R5893.

[3] RUDD R. E. and BROUGHTON J. Q., *Phys. Rev. B*, **72** (2005) 144104.

[4] KOBAYASHI R., NAKAMURA T. and OGATA S., *Mater. Trans.*, **49** (2008) 2541.

[5] YANG S. and QU J., *Modelling Simul. Mater. Sci. Eng.*, **22** (2014) 065011.

[6] STEVENS M. J., *Macromolecules*, **34** (2001) 2710.

[7] XIONG L., TUCKER G., MCDOWELL D. L. and CHEN Y., *J. Mech. Phys. Solids*, **59** (2011) 160.

[8] KAJITA S., WASHIZU H. and OHMORI T., *Euro. Phys. Lett.*, **87** (2009) 66002.

[9] KAJITA S., WASHIZU H. and OHMORI T., *Phys. Rev. B*, **82** (2010) 115424.

[10] KAJITA S., WASHIZU H. and OHMORI T., *Phys. Rev. B*, **86** (2012) 07545.

[11] WASHIZU H., KAJITA S., TOHYAMA M., OHMORI T., NISHINO N., TERANISHI H. and SUZUKI A., *Faraday Disc.*, **156** (2012) 279.

[12] CHEN M. Y., HONG Z. H., FANG T. H., KANG S. H. and KUO L. M., *T. Can. Soc. Mech. Eng.*, **37** (2013) 927.

[13] SHARMA G., LIMAYE P. K., RAMANUJAN R. V., SUNDARARAMAN M. and PRABHU N., *Maters. Sci. Eng. A*, **386** (2004) 408.

[14] TABAN E., GOULD J. E. and LIPPOLD J. C., *Mater. Des.*, **31** (2010) 2305.

[15] GUERRA R., TARTAGLINO U., VANOSSI A. and TOSATTI E., *Nat. Mater.*, **9** (2010) 634.

[16] LIU Y. L., GREY F. and ZHENG Q. S., *Sci. Rep.*, **4** (2014) 4875.

[17] KAJITA S., *Phys. Rev. B*, **94** (2016) 033301.

[18] RAYNE J. A., *Phys. Rev.*, **122** (1961) 1714.

[19] PENSINI E., SLEEP B. E., YIP C. M. and O'CARROLL D., *Colloids and Surfaces A: Physicochem. Eng. Aspects*, **433** (2013) 104.

[20] IWASAWA M., TANIKAWA A., HOSONO N., NITADORI K., MURANUSHI T. and MAKINO J., *Publ. Astron. Soc. Japan*, **68** (2016) 54.

[21] SANG Y., DUBÉ M. and GRANT M., *Phys. Rev. E*, **77** (2008) 036123.

[22] ISRAELACHVILI J. N., *Intermolecular and surface forces* (Elsevier, California) 2011.

[23] MORITA S., FUJISAWA S. and SUGAWARA Y., *Surf. Sci. Rep.*, **23** (1996) 1.

[24] CARPICK R. W., FLATER E. E., SRIDHARAN K., OGLETREE D. F. and SALMERON M., *JOM*, **56** (2004) 48.

[25] CARPICK R. W. and SALMERON M., *Chem. Rev.*, **97** (1997) 1163.

[26] SZLUFARSKA I., CHANDROSS M. and CARPICK R. W., *J. Phys. D: Appl. Phys.*, **41** (2008) 123001.

[27] KUZNETSOV V. D., *Metal Transfer and Build-up in Friction and Cutting* (Pergamon Press, London) 1966.

[28] FURLAN K. P., PRATES P. B., SANTOS T. A. D., DIAS M. V. G., FERREIRA H. T., NETO J. B. R. and KLEIN A. N., *J. Alloy Comp.*, **652** (2015) 450.

[29] BLAU P. J., *Friction Science and Technology: From Concepts to Applications* (CRC Press, New York) 2009.

[30] LEMUL H. G. and TRZEPIECIŃSKI T., *J. Mech. Eng.*, **59** (2013) 41.

[31] MAVEYRAUD C., BENZ W., SORNETTE A. and SORNETTE D., *J. Geophys. Res.,* **104** (1999) 28769.

[32] BHUSHAN B., *Tribology and Mechanics of Magnetic Storage Devices* (Springer Verleg, New York) 1996.

[33] LIU E., BLANPAIN B., CELIS J. P. and ROOS J. R., *J. Appl. Phys.*, **84** (1998) 4859.

[34] CHAN S. M. T., NEU C. P., KOMVOPOULOS K. and REDDI A. H., *J. Biomech.*, **44** (2011) 1340.

[35] CHOWDHURY M. A., NURUZZAMAN D. M., MIA A. H. and RAHAMAN M. L., *Tribology in Industry*, **34** (2012) 18.

[36] KATANO Y., NAKANO K., OTSUKI M. and MATSUKAWA H., Sci. Rep., **4** (2014) 6324.

[37] ALARCÓN H., SALEZ T., POULARD C., BLOCH J. F., RAPHAËL E., VERESS K. D. and RESTAGNO F., *Phys. Rev. Lett.*, **116** (2016) 015502.

**Figure captions:**

**Fig. 1:** The CG process of the atomistic system into the simulated particle ones. (a) A bcc atomic Fe cubic one of 8 unit cells. (b) A sc unit cell of the particle system is taken from the first coarse-graining in which one particle is yielded by grouping an atomic unit cell. (c) A sc unit cell of the particle system is taken from the final coarse-graining in which one particle is yielded by grouping $N_l^3$ unit cells of the previous CG system. (d) The simulation system of 432000 particles at the starting time of the sliding.

**Fig. 2:** Sliding time-dependence of the friction force, normal force and friction coefficient of the systems, left column for the 16000 particles system and right column for the 432000 particles system.

**Fig. 3:** Sliding time-dependence of the friction coefficient of the 16000 particles system. The simulation conditions of load of 0.1 μN, $k_z = k_y = 0.1 k_x$ with (a) $k_x = K$ and (b) $k_x = 0.1K$.

**Fig. 4:** Dependence of the friction coefficient of the systems on the ratio of the spring constant $k_z / k_x$.

**Fig. 5:** Dependence of the friction coefficient of the systems on the externally applied normal load.

**Fig. 1:**

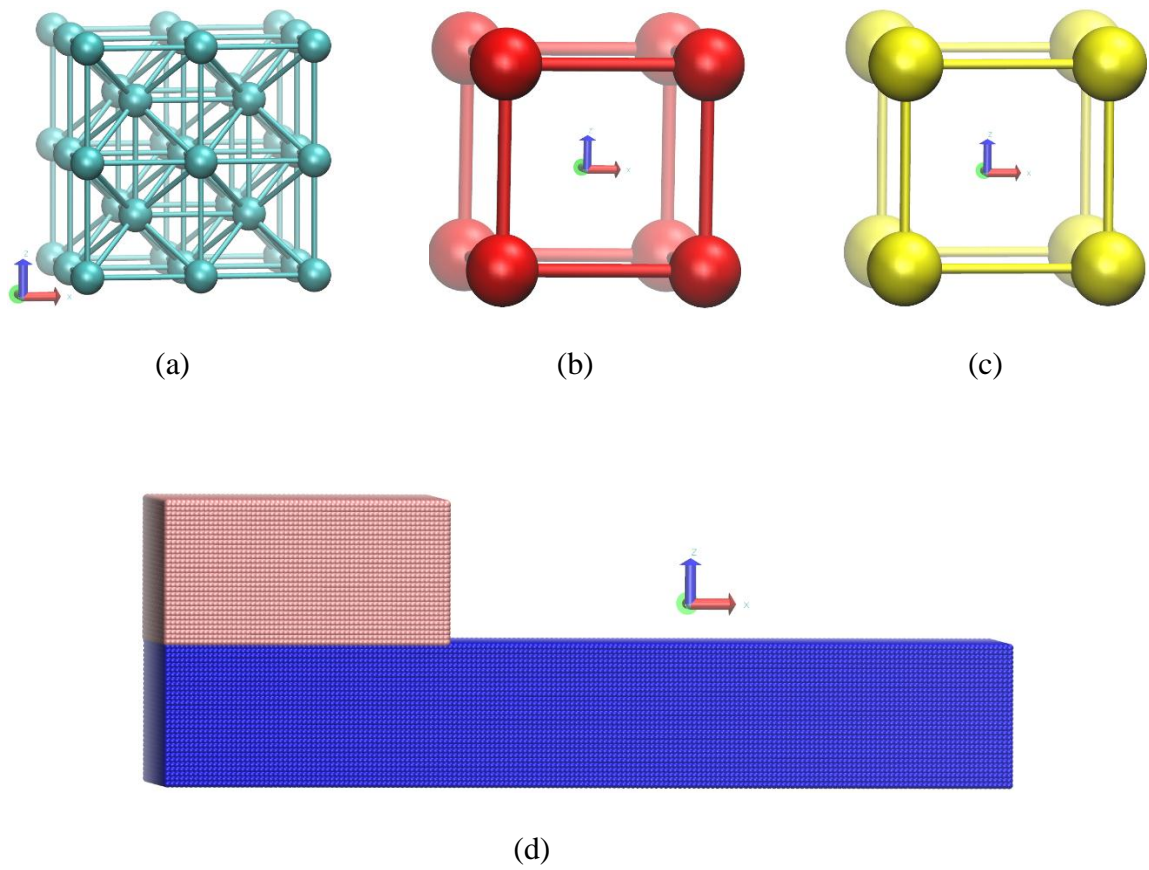

(a)          (b)          (c)

(d)

**Fig. 2:**

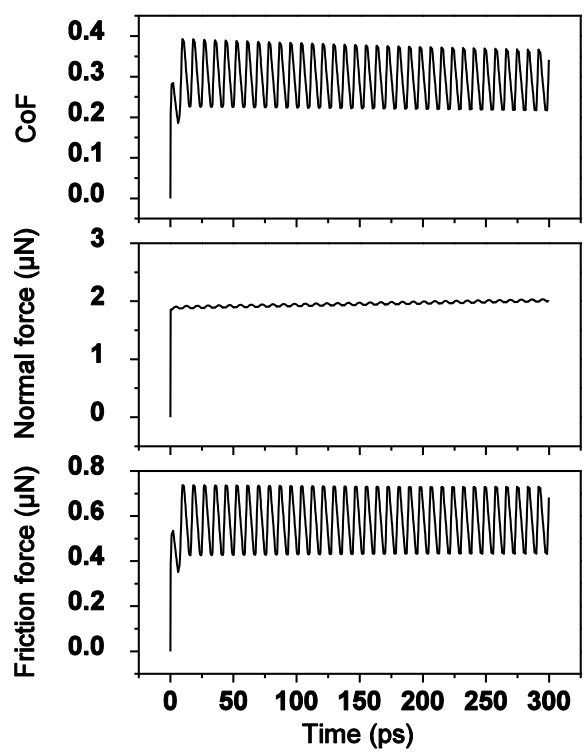 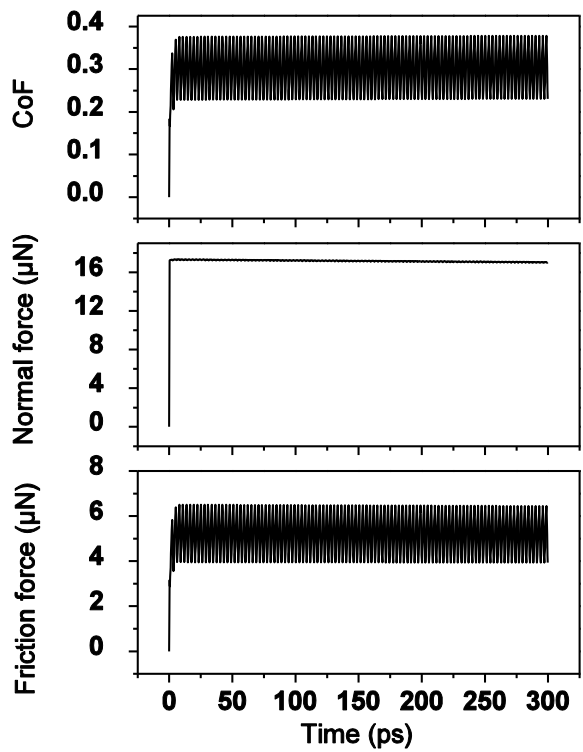

Fig. 3:

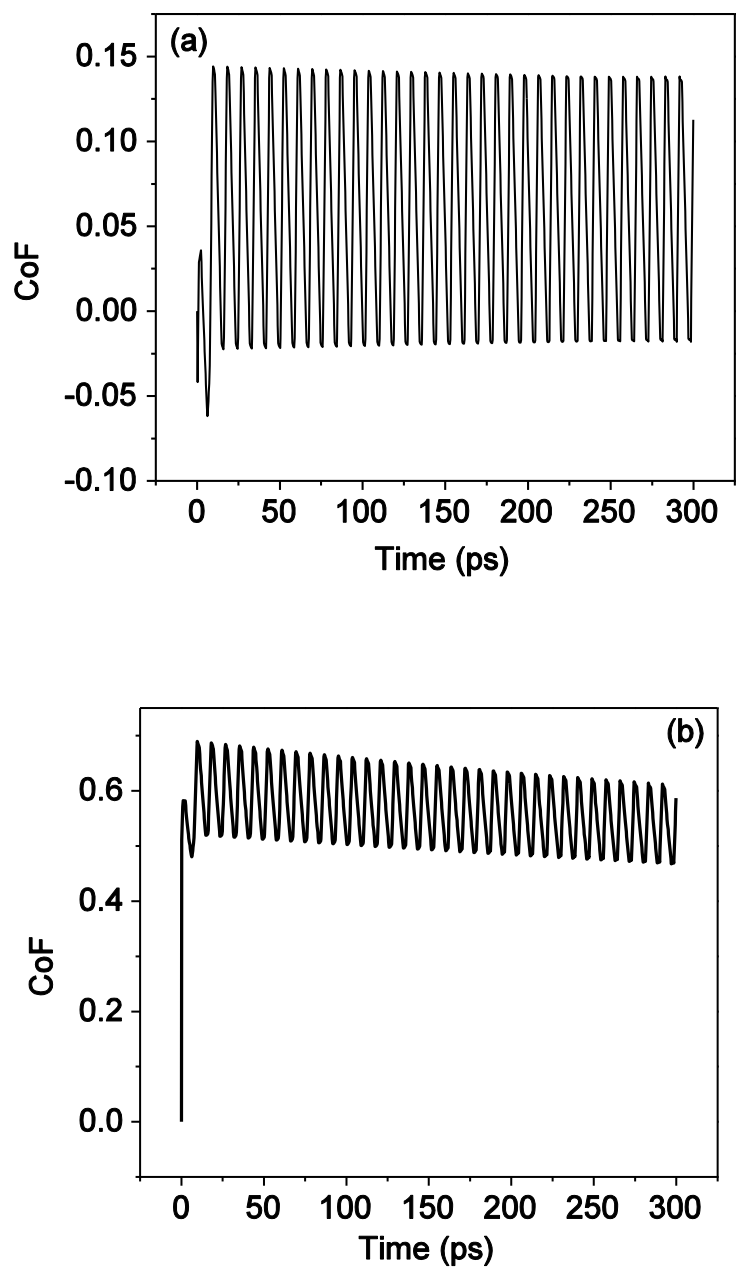

**Fig. 4:**

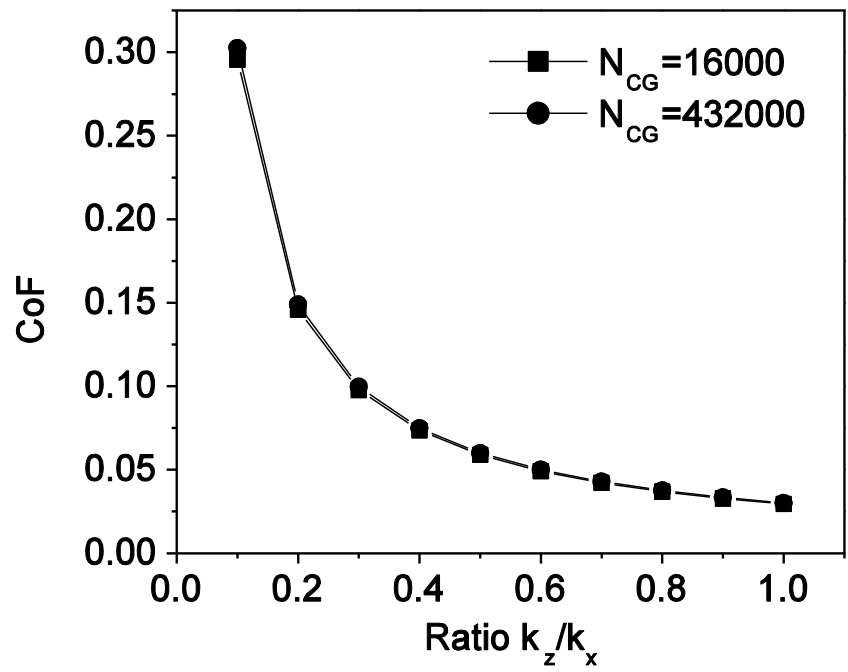

**Fig. 5:**

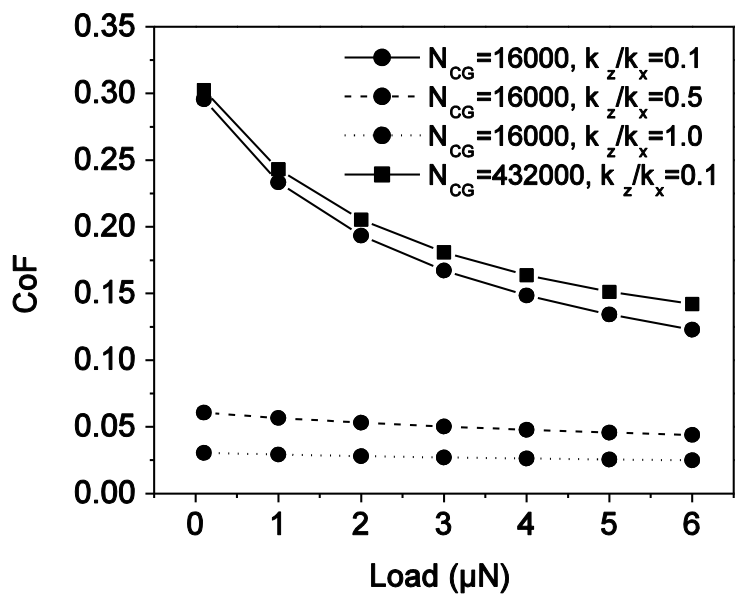